%% file: paper.tex
\documentclass[conference,a4paper]{IEEEtran}
\IEEEoverridecommandlockouts
\usepackage{cite}
\usepackage{amsmath,amssymb,amsfonts}
\usepackage{algorithmic}
\usepackage{graphicx}
\usepackage{textcomp}
\usepackage{xcolor}
\usepackage{csquotes}
\usepackage{booktabs}
\usepackage{tabularx}
\usepackage{url} 
\usepackage[inline,shortlabels]{enumitem}

\usepackage{tikz}
\newcommand*\circled[1]{\tikz[baseline=(char.base)]{
    \node[shape=circle,draw,inner sep=1pt] (char) {#1};}}

\newcommand{\myissue}[2]{\underline{#2}}

\usepackage{siunitx}
\usepackage[nolist]{acronym} 
\acused{IoT}
\acused{LoRaWAN}
\acused{GPS}

\newcommand*{\myOpenSource}{open source }

\usepackage{xcolor,colortbl}

\begin{document}
\bstctlcite{IEEEexample:BSTcontrol}
\title{Implementation of OpenAPI Wireshark Dissectors to Validate SBI Messages of 5G Core Networks\\
}

\author{
	\IEEEauthorblockN{
		Lukas Schauer\IEEEauthorrefmark{1},
		Thorsten Horstmann\IEEEauthorrefmark{1}\IEEEauthorrefmark{2}, 
		Steffen Druesedow\IEEEauthorrefmark{3}, 
		Michael Rademacher\IEEEauthorrefmark{1}\IEEEauthorrefmark{2}
	}
	\IEEEauthorblockA{\IEEEauthorrefmark{1}Hochschule Bonn Rhein-Sieg, Sankt Augustin, Germany, {firstname.lastname}@h-brs.de}
	\IEEEauthorblockA{\IEEEauthorrefmark{2}Fraunhofer FKIE, Bonn, Germany, {firstname.lastname}@fkie.fraunhofer.de}
	\IEEEauthorblockA{\IEEEauthorrefmark{3}Telekom Deutschland GmbH, Berlin, Germany, {firstname.lastname}@telekom.de}
}

\maketitle

\begin{abstract}
This paper introduces a novel Wireshark dissector designed to facilitate the analysis of \ac{SBI} communication in 5G Core Networks.
Our approach involves parsing the OpenAPI schemes provided by the 5G specification to automatically generate the dissector code.
Our tool enables the validation of 5G Core Network traces to ensure compliance with the specifications. 
Through testing against three open-source 5G Core Network projects, we identified several issues where messages deviate from specification standards, highlighting the significance of our implementation in ensuring protocol conformity and network reliability.
\end{abstract}


\input{sections/introduction}
\input{sections/background}
\input{sections/related_work}
\input{sections/concept}
\input{sections/evaluation}
\input{sections/summary}

\input{acro}
\bibliographystyle{IEEEtran}
\bibliography{./bibabbrv.bib, ./lit.bib}
\end{document}

%% file: sections/introduction.tex
\section{Introduction}
Mobile networks based on 5G have become a reality and different vendors provide commercial equipment for \acp{UE}, \acp{RAN} and the \ac{5GC}. Compared to previous generations of mobile networks, interoperability between hardware and software components from different vendors has become even more crucial. This demand is particularly evident due to the heavy usage of concepts like \ac{SDN} and \ac{NFV} and the resulting microservices, which on the one hand lead to agility and cost-efficiency, but on the other hand to strict requirements towards the specification of the used protocols~\cite{moreira2020next}. In this work, we put these strict requirements to test because despite the fact that 5G is gaining more and more popularity, there is one thing nobody seems to care about: protocol validity.

There are reasons why protocol validity is a more crucial factor in 5G than it was for previous generations of mobile networks. 
The main driver for that change was a completely new concept for mobile core networks, away from vendor specific binary protocols and monolithic systems towards a well accepted and commonly known technology based on microservices and RESTful protocols based on OpenAPI.
This change led to a large increase of different \ac{5GC} implementations, not just in the open-source communities - also new commercial players entered the market.
Such a variety comes with advantages for \acp{MNO} but also with challenges, especially since the \ac{3GPP} specifications are rather complex and sometimes leave space for interpretations.
Ensuring that components from different vendors interoperate correctly with each other, but also with available emulators and testing environment can be a tedious task.
It often means tough study of the specifications and manual validation of large traces, still leaving the human factor as final error source.
Objective tooling for validation not only helps operators to ensure correct operation, it also helps implementers during the development cycle and eventually may contribute to a cleaner \ac{5GC} ecosystem overall.

The motivation for this work emerges from our practical experiments with different \ac{5GC}: \textbf{We found various inconsistencies and issues with different implementations for the 5G \ac{SBA}}. To further analyze these inconsistencies and issues, we started looking for tooling. However, to our surprise, we were unable to identify any. Specific tooling was seemingly available within the paid products of various vendors, but they obviously targeted only big \ac{MNO}.

In the end we looked at our options: The 5G specifications contain ASN.1 definitions for the binary part of core network communication, which basically can already be validated by Wireshark in various ways, but they also contain OpenAPI specifications, which describe the HTTP/2 API that is being used as the primary way of communication within the \ac{SBA}. We quickly found that OpenAPI is limited to the validation of single request-response-pairs. However, this also provides a chance for a fairly straight-forward process, which we present in this work. In particular, our contributions are the following: 
\begin{enumerate}[wide]
\item \textbf{We describe our implementation of a validator for OpenAPI specifications.} We implemented this validator as a Wireshark dissector, since Wireshark was already being used by us and the scientific community in general.
\item \textbf{We used our validator to analyze a simple scenario} using various \myOpenSource implementations of \ac{5GC} and identified various issues within these implementations.
\item Most importantly, \textbf{we published our implementation \myOpenSource\cite{telekomO2:online}}.
\end{enumerate}

The rest of this work is structured as follows: In Section~\ref{sec:background} we briefly provide the required background regarding \mbox{OpenAPI}, \ac{5GC} and Wireshark. Section~\ref{sec:related_work} discusses different tools which are related to this work. The main part of this work is Section~\ref{sec:concept}, where we describe our concept and implementation. To demonstrate the usage of our tool, Section~\ref{sec:evaluation} reports on the evaluation of three widely-used \ac{5GC} implementations. We provide a summary and future work in Section~\ref{sec:summary}.

%% file: sections/background.tex
\section{Background}\label{sec:background}
The \textbf{OpenAPI} specification not only defines the API endpoints and their associated operations but also provides a structured way to describe the request and response data models~\cite{OpenAPI:online}.
This includes specifying the expected data formats, field types, and validation rules for the message payloads exchanged between \acp{NF}.
For example, the data model for a Session Establishment Request sent to the \ac{SMF} may include fields like the \ac{SUPI} of type \enquote{string} with a regular expression pattern, and nested objects.
Well-defined message structures are crucial in the \ac{5GC}, as they ensure consistent interpretation of the data across different network elements, facilitate automated validation, and reduce the likelihood of interoperability issues.

The 5G mobile networks have brought about a paradigm shift in the architecture and design of cellular core networks.
One of the key innovations in 5G is the adoption of a \ac{SBA}, where network functions are decomposed into modular services that communicate with each other through well-defined APIs \cite{10258072}.
This approach, known as the \acf{SBI}, promotes flexibility, scalability, and efficient resource utilization within the \ac{5GC}. The \ac{SBI} in the \ac{5GC} comprises a set of web-based APIs that enable communication between various \acp{NF}, such as the \ac{AMF}, \ac{AUSF}, \ac{UDM}, and others. 
These APIs are defined using the OpenAPI Specification and follow the RESTful architectural style, employing standard HTTP methods (GET, POST, PUT, DELETE) and JSON data formats for request and response payloads~\cite{ts-129-501}.

\textbf{Wireshark}, the widely-used network protocol analyzer, plays an important role in troubleshooting and validating network communications. As a productive packet capturing and analysis tool, Wireshark provides a huge set of dissectors for nearly all common network protocols, to decode and inspect network packets and flows \cite{Wireshark}.
To enhance its protocol dissection capabilities, Wireshark can be extended through custom dissectors written in the Lua programming language~\cite{Lua}. These dissectors allow developers to define rules for parsing and displaying protocol data structures without the need to compile the code.
Furthermore, Wireshark's ability to apply display filters, follow packet streams, and export captured data for further analysis makes it a solid tool for validating the correct implementation of the \ac{5GC} \ac{SBA}.

%% file: sections/related_work.tex
\section{Related Work}\label{sec:related_work}
Validation of RESTful APIs has been comprehensively discussed in \cite{karlsson2020quickrest}. In their work, the authors propose the tool \textbf{QuickREST} which is built up-on the concept of property-based testing (PBT). The idea of PBT is to automatically generate input data and evaluate if the API reacts to this input according to the specification. This approach is significantly different from our concept. In this work, we passively monitor traffic and conduct the validation offline.

The \textbf{5G Trace Visualizer} by Deutsche Telekom is an \myOpenSource toolkit that allows to visualize traces from a 5G core network \cite{5g-trace-visualizer}. It uses Wireshark in the background to decode the traces and combines its output with metadata from Kubernetes and/or OpenStack. The resulting data can be visualized using an \myOpenSource UML diagram tool to generate message sequence charts or by using a Python based plotting library to draw various metrics. The Visualizer itself is scripted using iPython/Jupyter notebooks and contains various example scripts, e.g., for activity in the core network correlated to resource usage in a Kubernetes cluster. Unfortunately, no actual validation of data according to the OpenAPI descriptions is implemented.

\textbf{APIClarity} by Cisco is a tool designed for analyzing and validating APIs described using OpenAPI specifications. It offers a wide range of features, including automatic generation of OpenAPI documentation based on real traffic on existing APIs and fuzzing of API endpoints \cite{APIClarity}. APIClarity is mainly designed to be used as proxy between an API and the rest of the world, but it also offers support for network taps that can passively capture API traffic. Unfortunately, APIClarity seems to be unsuited for the analyzing and validating traffic of \ac{5GC}, as it was designed primarily for small and simple APIs. Our attempts to import parts of the 5G OpenAPI documents were not successful. 

%% file: sections/concept.tex
\section{Concept and Implementation}\label{sec:concept}

\begin{figure*}[h]
  \centering	
  \includegraphics[width=\textwidth]{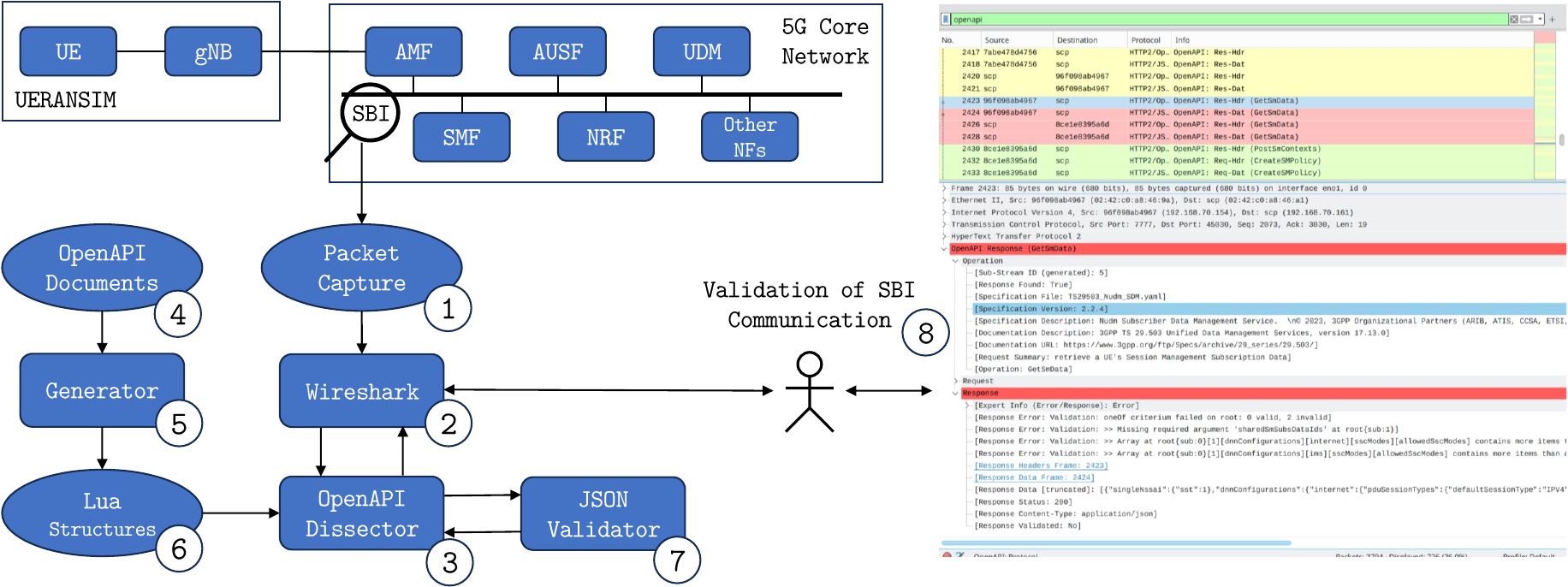}
  \caption{Architecture of the implemented dissector and the evaluation scenario.}
  \label{fig:architecture}
\end{figure*}

Our approach is based on several software components interacting with each other. Figure~\ref{fig:architecture} provides an overview of the plugin architecture. In the following, we will highlight several parts. We enumerated these parts in Figure~\ref{fig:architecture} for
a better readability throughout this section. 

We decided to write a Wireshark \circled{2} plugin capable of parsing \ac{SBI} messages, allowing for easy validation of specification conformity of core network components. The plugin itself is written in Lua, which makes it extremely portable. Only a few files need to be placed in the Wireshark plugins directory and it simply starts working, independent of CPU architecture, operating system or other user specific parameters.

The plugin is implemented as a post-dissector. This approach allows us to easily build upon the existing HTTP/2 dissection. Unfortunately, this means that currently no dissection of live-traffic is possible, all input data needs to have been captured previously, so a PCAP file \circled{1} is used as input.

The main plugin \circled{3} begins by trying to find corresponding request and response headers as well as data. It uses Wiresharks TCP flow numbers and HTTP/2s internal stream IDs to find matching packets. For each pair, the path from the request headers is used to look up the corresponding part of the OpenAPI documents.

The 3GPP publishes and frequently updates their OpenAPI documents in a publicly available repository in the form of various yaml files~\cite{3GPPForge}. We use these yaml files \circled{4} as the ground truth for our tool. In theory, it would be possible to just put the corresponding yaml files in a directory for parsing by the plugin, but initial tests showed that this would take quite a considerable amount of time to find the corresponding specification for each request. To avoid having to wait for this work to be done on every launch of the dissector plugin, a Python based generator script \circled{5} reorganizes path and component references in an easy-to-parse way. The script mainly creates a nested data structure \circled{6}, allowing easy access to required schema data indexed by request paths. These paths do not purely consist of static strings but allow for variables of various data types. The generator creates corresponding regular expressions for these complex path specifications. Thanks to the inclusion of the optimized regular expression library PCRE2~\cite{pcre2:online} inside of Wiresharks Lua interpreter, the patterns can quickly be resolved during runtime. For each request, the dissector searches this nested structure and provides the found schema data to the validation script.

When the connection establishment has not been captured (due to starting a capture in an already running core network) some headers might get lost due to HTTP/2s built-in compression. Our plugin tries to augment headers with best-effort estimations where possible. Some headers, e.g. Content-Type, are generated by looking at the format of the actual content, other headers are chosen from the OpenAPI documents.

After finding a matching path specification in the documents all headers, data and the path specifications get passed over to our validation script \circled{7}, which parses the provided JSON data and checks the given path specifications against it. For simple elements like strings and numbers the checks are rather simple. They basically consist of a list of conditions that might need to be matched (e.g. number in certain range, string matching a pattern, etc.) and return any found validation errors.

More complex elements like \enquote{objects} and \enquote{arrays} which contain other elements needed a bit more work, especially since the documentation of OpenAPI itself is very vague about some more specific behavior, e.g. when using negation. We approached this by implementing the behavior of our validation close to the corresponding sections we found in the 5G specifications OpenAPI documents.

The biggest challenge was adding support for \texttt{oneOf} / \texttt{anyOf} / \texttt{allOf} properties. They allow for an
element at a specific location to be validated in various ways. They all take a list of various schemas which need to be validated against an element. The names imply how many of these schemas are allowed to validate for the given element. The \texttt{oneOf} property is special, because it allows for the additional definition of a discriminator value, which is then used to decide which of the given schemas the element gets validated against.

If any subelement validation fails the objects or arrays validation fails as well, so any validation errors are propagated back down to the JSON root element. Validation information and metadata is then added to the packet view \circled{8} in Wireshark. With a simple click on a packet a user can see what document it was validated with, what operation it belongs to, have links to related packets (request/response, notifications/subscriptions), and especially see if it was detected as valid according to the OpenAPI document or if there have been any issues during validation.

%% file: sections/evaluation.tex
\section{Evaluation}\label{sec:evaluation}
To test our implementation we decided to evaluate current versions of three different open-source 5G Core Network projects: Open5GS (v2.7.0)~\cite{open5GS:online}, free5GC (v3.3.0)~\cite{free5GC:online} and OpenAirInterface (v2.0.1)~\cite{OpenAirI88:online}.

For this evaluation, we set up simple test networks, each with one of the available implementations and an instance of \mbox{UERANSIM}. \mbox{UERANSIM} is an open-source project that provides a software implementation of the \ac{RAN} as well as the \ac{UE} side of the 5G network~\cite{UERANSIM}.
It is designed to interoperate with any 5GC to enable end-to-end testing and validation of 5G network scenarios. To this end, it simulates the behavior of a gNodeB (gNB) and a \ac{UE}, including the establishment of radio bearers, registration procedures, and exchanging control plane and user plane data with the \ac{5GC}.
By omitting the actual radio part, UERANSIM allows setting up functional 5G networks without the need for physical \ac{UE} or gNB devices. Our test case for this work is the initialization of the 5G network, followed by a \ac{UE} registration with corresponding \ac{PDU} session establishment and ending with the simulated UEs deregistration. 
For each of the three evaluated \ac{5GC} projects, we performed this scenario and observed issues that our validator reported.
Table~\ref{tab:overview-results}~summarizes our results. In the following subsection, we provide detailed descriptions of the issues we found in the different implementations. This description should emphasize the complex nature of \ac{SBI} messages and why a tool like the one presented in this work provides real benefit.

\begin{table*}[bth!]
\caption{Overview about the evaluation results. Five selected key issues identified per open-source \ac{5GC} implementations.}
\label{tab:overview-results}
\begin{tabularx}{\textwidth} { 
   p{2cm}
   >{\raggedright\arraybackslash}X 
   >{\raggedright\arraybackslash}p{2.2cm}
   >{\raggedright\arraybackslash}X
   >{\raggedright\arraybackslash}X
   >{\raggedright\arraybackslash}X}
\toprule
Implementation & \multicolumn{5}{c}{Issues} \\ \midrule
Open5GS & Service list in NF registration & Wrong version of UDR API & Creation of NRF subscriptions & Missing address information in resp. to NF registration & Handling of UEs session management subscription data \\ \midrule
free5GC & Invalid content-type for UEAuthenticationCtx & Wrong version of UDR and SDM API & Invalid ConfirmAuth messages & Nulled content during SearchNFInstances call & Invalid age values in UpdateSmContext messages \\ \midrule
OpenAirInterface & Invalid NF registration & Wrong version of UDR API & Invalid NRF subscription condition & Invalid ConfirmAuth messages & Invalid SD encoding \\ \bottomrule
\end{tabularx}
\end{table*}

\subsection{Open5GS}
Open5GS is a prominent open-source implementation of the 5G Core Network, providing both 5G Standalone and Non-Standalone \acp{NF}.
Developed primarily in the C programming language by a community of contributors, it is used by researchers and developers to experiment with and test 5G Core features and capabilities in a wide range of projects. During our tests, we found five notable issues regarding the OpenAPI specification correctness.

\myissue{1}{Service list in NF registration:} During startup, some calls to the \ac{NRF} contain an empty \texttt{nfServiceList} during the \texttt{RegisterNFInstance} operation. According to the 3GPP Technical Specification (TS) 29.510, describing the NRF, the \texttt{nfServiceList} must contain at least one item (if present)~\cite{3gpp.29.510}.

\myissue{2}{Wrong version of UDR API:} According to Release 17 of the 5G specification the UDR API should be v2~\cite{3gpp.29.504}, but Open5GS still uses v1, which last seems to have been used in early versions of Release 15. We didn't further validate the correctness of these messages, as our validator currently has no support for multiple specification versions at the same time. Also the specifications for these early releases are not easily available since they are only included in the human-readable documentation of the 5G specification and are provided for information purposes only.

\myissue{3}{Creation of NRF subscriptions:} A \texttt{subscriptionId} is generated by the \ac{NRF} in the \texttt{CreateSubscription} messages, which identifies the corresponding subscription. The according NRF specification~\cite{3gpp.29.510} describes this identifier to consist of the \ac{MCC} and \ac{MNC}, followed by a generated part, but the ID used by Open5GS consists purely of a UUID.

\myissue{4}{Missing address information in resp. to NF registration:} According to TS 29.510~\cite{3gpp.29.510} the response during an \texttt{RegisterNFInstance} operation should be an \texttt{NFProfile} description, which needs to contain at least one piece of addressing information (a fully qualified domain name, an IPv4 or an IPv6 adress), but the Open5GS NRF does not include this information in the response.

\myissue{5}{Handling of UEs session management subscription data:}
During the response to a \texttt{GetSmData} call the Open5GS \ac{UDM} sends a \texttt{SessionManagementSubscriptionData} packet containing a \texttt{DnnConfiguration}.
Inside the configuration \texttt{allowedSscModes} are defined for the IP Multimedia Subsystem (IMS) and the Internet service. According to the UDM specification, this list should contain a maximum of two entries~\cite{3gpp.29.503}, but in our example setup the Open5GS \ac{UDM} returns a list containing three items.

\subsection{free5GC}
Free5GC is another notable actively developed open-source 5G Core Network implementation using the Go programming language. Free5GC’s original goal was to provide academics with a platform to test and prototype 5G systems. 
Its feature set and open-source nature not only facilitate research and testing but also offer commercial value, particularly for deploying private 5G networks.

\myissue{1}{Invalid content-type for UEAuthenticationCtx:} During POST requests containing \texttt{UEAuthenticationCtx} messages the content-type header is set to \texttt{application/json}, but the \ac{AUSF} specification describes these messages as \texttt{application/3gppHal+json}~\cite{3gpp.29.509}. While this probably would not cause any issues as it is the only expected data structure in this case, it still remains at least a minor violation of the provided specifications. The validator has a workaround for these special cases, where it basically ignores the given content type and looks up a content type from the specification. This only works for requests and responses with a single available content type, but it is better than simply ignoring the content of affected packets.

\myissue{2}{Wrong version of UDR and SDM API:} Unfortunately, free5GC does not seem to specify the minor version of the used 5G specification version, so we assume the latest version of Release 15. Similar to Open5GS, it is still using UDR and various other APIs in v1, which are no longer in use since later versions of Release 15. As mentioned before, we have no machine-readable specifications for these older API versions and were unable to validate these messages accordingly.

\myissue{3}{Invalid ConfirmAuth messages:} Instead of a valid \texttt{AuthEvent} message the response to \texttt{ConfirmAuth} messages is simply the value \texttt{null}. It also does not contain the expected location header to a created resource~\cite{3gpp.29.504}. While the return data might not be important, especially with a correctly provided status code, this might still result in issues when used with different core network components.

\myissue{4}{Nulled content during SearchNFInstances call:} During the response to a call to \texttt{SearchNFInstances} the \texttt{nfInstances} parameter is simply set to \texttt{null}. According to the specification~\cite{3gpp.29.510}, this should always be an array, but it might be empty if no instances have been found.

\myissue{5}{Invalid age values in UpdateSmContext messages:} Inside the \texttt{UpdateSmContext} message an \texttt{ageOfLocationInformation} variable is set to a negative number, but according to the common data types specification for the \ac{SBI}~\cite{3gpp.29.571}, this value needs to be in the range between 0 and 32767.

\subsection{OpenAirInterface}
\ac{OAI} is an open-source project consisting of both a new 5G Radio Access Network (RAN) and a Core Network implementation, running on general purpose x86 computing hardware and commercial off-the-shelf Software Defined Radio (SDR) cards.
The code is developed by a world-wide industrial and academic community, governed under the OAI Software Alliance consortium.

\myissue{1}{Invalid NF registration:} Similar to Open5GS, the most calls to the NRF contain an empty \texttt{nfServiceList} during the \texttt{RegisterNFInstance} operation, while they must contain at least one item (if present). In addition, the \texttt{SupiRange} type used by some operations is not encoded correctly. It contains the \texttt{pattern} attribute as well as \texttt{start} and \texttt{end} fields, which is not allowed. The specified values are also not in accordance with the specification~\cite{3gpp.29.510}.

\myissue{2}{Wrong version of UDR API:} Similar to Open5GS and free5GC, \ac{OAI} uses \texttt{v1} as endpoint API version for the \texttt{authentication-status} operation, while the specification requires \texttt{v2}.
This shows a widespread misinterpretation of the specification across multiple projects.

\myissue{3}{Invalid NRF subscription condition:} In the request to create a subscription to a specific \ac{NF} instance in the NRF, the \ac{NF} is determined according to different criteria specified by the \texttt{subscrCond} attribute of the \texttt{SubscriptionData} object type. This attribute is a composed schema that selects only one of the listed schemas under the keyword \texttt{oneOf}. Apparently, the encoding is implemented incorrectly by \ac{OAI}. An additional mapping from \texttt{subscrCond} to \texttt{nfTypeCond} is inserted in the JSON structure to encode the \texttt{oneOf} concept.
However, this does not match the expected OpenAPI specification~\cite{3gpp.29.510}, since the value of \texttt{subscrCond} becomes the composition of two objects rather than a single object. The issue is a consequence of a serious abandoned bug in the library \enquote{OpenAPI Generator} used by \ac{OAI} for the OpenAPI client types de- and encoding. It is interesting to note, that the exact same problem was found in Ericsson’s Network Core Test System by Davide Donato~\cite{openAPIerlang}. This emphasizes the complexity of the rules for encoding OpenAPI concepts, which makes manually writing code sensitive to inconsistencies and bugs.

\myissue{4}{Invalid ConfirmAuth messages:} During the authentication process of an \ac{UE}, the \ac{AUSF} responses to the \ac{AMF} the result of the \texttt{5g-aka-confirmation} operation as an \texttt{AuthResult} enumeration type~\cite{3gpp.29.504}. The \ac{OAI} implementation of the \ac{AUSF} is using a simple boolean type instead of the specified enumeration values. This can lead to serious security problems if other \ac{AMF} implementations are used that are unable to handle this incorrect encoding appropriately.

\myissue{5}{Invalid \ac{SD} encoding:} In several messages the \ac{SD} is encoded incorrectly. The specification requires an exact 6-digit number to distinctive 5G network slices if they use the same Slice/Service Type (SST). However, \ac{OAI} encodes the \ac{SD} with variable length, which results in a minor specification violation~\cite{3gpp.29.571}.

%% file: sections/summary.tex
\section{Summary and Future Work}\label{sec:summary}
This work presents our prototype for an OpenAPI Wireshark dissector to validate \ac{SBI} messages of a \ac{5GC}. The implementation aspects are presented in detail in this work. In addition, we use a simple scenario to analyze messages from three different open-source \ac{5GC} implementations. Despite the simplicity of the scenario, we found various issues where the implementation differs from the protocol specification.

With the implementation of this tool, we want to contribute to a better implementation of \aclp{5GC}. Additionally, we provide users, for example experts working at \acl{MNO} or researchers, the possibility to better debug their network infrastructure. Lastly, if protocol validity is improved overall, this directly leads to more and better interoperability in the 5G ecosystem.  

Although our prototype is working as expected and fully functional, we have already identified several aspects which can be improved. In the current state, certain components of the OpenAPI specification are implemented solely as minimal placeholders.
Guessing the missing HTTP/2 header fields, resulting from stream compression, is based on a simple approach, which may lead to the selection of incorrect specifications for validation on rare occasions.
Our main goal for future work is to add the possibility for validation of longer procedures, that consist of more than a simple request-response-pair. Our basic idea is to match against an ordered list of required operations to see if e.g.\ a \ac{NF} registration and deregistration is done properly. This could be extended by matching service IDs, allowing for validation of multiple parallel sequences of operations, potentially also involving \acp{UE} in later steps.

%% file: acro.tex
\begin{acronym}[]
\acro{3GPP}{3rd Generation Partnership Project}
\acro{SBI}{Service-Based Interface}
\acro{SBA}{Service-Based Architecture}
\acro{NF}{Network Function}
\acro{AMF}{Access \& Mobility Management Function}
\acro{AUSF}{Authentication Server Function}
\acro{NRF}{Network Repository Function}
\acro{SMF}{Session Management Function}
\acro{UDM}{Unified Data Management}
\acro{MNO}{Mobile Network Operator}
\acro{SUPI}{Subscription Permanent Identifier}
\acro{5GC}{5G Core Network}
\acro{OAI}{OpenAirInterface}
\acro{SD}{Slice Differentiator}
\acro{UE}{User Equipment}
\acro{RAN}{Radio Access Network}
\acro{SDN}{Software-Defined Networking}
\acro{NFV}{Network Functions Virtualization}
\acro{MCC}{Mobile Country Code}
\acro{MNC}{Mobile Network Code}
\acro{RAN}{Radio Access Network}
\acro{PDU}{Packet Data Unit}
\end{acronym}